\documentclass{article}
\usepackage{spconf,amsmath,graphicx}
\usepackage{multirow}
\usepackage{enumerate}
\usepackage{booktabs}
\usepackage[colorlinks,linkcolor=black,anchorcolor=black,citecolor=black]{hyperref}
\usepackage{setspace}
\usepackage[mathscr]{eucal}
\usepackage{amssymb}

\title{Cross-Modal Audio-Visual Co-learning for Text-independent Speaker Verification}
\name{Meng Liu$^{1,2}$, Kong Aik Lee$^2$, Longbiao Wang$^1$, Hanyi Zhang$^{1}$, Chang Zeng$^{3}$, Jianwu Dang$^1$}

\address{$^1$College of Intelligence and Computing, Tianjin University, China \\
$^2$Institute for Infocomm Research, A$^\star$STAR, Singapore \\
$^3$National Institute of Informatics, Tokyo, Japan}

\begin{document}
\ninept
\maketitle
\begin{abstract}
Visual speech (i.e., lip motion) is highly related to auditory speech due to the co-occurrence and synchronization in speech production. This paper investigates this correlation and proposes a cross-modal speech co-learning paradigm. The primary motivation of our cross-modal co-learning method is modeling one modality aided by exploiting knowledge from another modality.  Specifically, two cross-modal boosters are introduced based on an audio-visual pseudo-siamese structure to learn the modality-transformed correlation. Inside each booster, a max-feature-map embedded Transformer variant is proposed for modality alignment and enhanced feature generation. The network is co-learned both from scratch and with pretrained models. Experimental results on the LRSLip3, GridLip, LomGridLip, and VoxLip datasets demonstrate that our proposed method achieves 60\% and 20\% average relative performance improvement over independently trained audio-only/visual-only and baseline fusion systems, respectively.
\end{abstract}

\section{Introduction}
Audio-visual lip biometrics \cite{luettin1996speaker} utilizing auditory speech (i.e. spoken utterances) and visual speech (i.e., lip motion)  has raised increasing attention recently \cite{aleksic2006audio}. Unlike visual biometrics using static face or iris images \cite{chung2018voxceleb2}, lip motion has a dynamic temporal behavior that could be aligned with auditory speech \cite{stafylakis2018deep}. When a person speaks, his/her voice, lip motion, and spoken words are three closely bound modalities of audio, visual, and text \cite{liu2013learning} and vary from utterance to utterance. Therefore, it is difficult to forge these identity characteristics at the same time \cite{yang2020preventing}. Due to these advantages, audio-visual lip biometrics could be applied to various mobile applications and highly secured financial identification systems.

Over the past decades, the development of lip biometrics has undergone a significant change from the classic machine-learning approaches to data-centric deep-learning models. In the former, support vector machine (SVM), hidden Markov model (HMM) \cite{luettin1996speaker}, and Gaussian mixture model (GMM) \cite{aleksic2006audio} were used in conjunction with appearance-based \cite{wang2012physiological} and shape-based \cite{faraj2007speaker} features derived from lip geometry. In the latter, lip image sequences were usually directly fed into deep neural networks \cite{liu2021deeplip}. In \cite{shi2022learning}, audio-visual speaker embedding was extracted from lip sequences with a pretrained AV-HuBERT model. However, these methods focused more on modality fusion but overlooked the correlation between auditory and visual speech \cite{zhang2020multimodal,sari2021multi,tao2020audio}.

In \cite{liu2021deeplip}, we presented the DeepLip system that fuses complementary information derived from auditory and visual speech. The visual stream uses a multi-stage convolutional neural network (MCNN) to extract visual speaker embedding. The audio stream employs an x-vector system \cite{snyder2018x} to extract audio speaker embedding. Although it achieves a satisfactory performance, the concurrency of auditory and visual speech is largely ignored in the preliminary work: no modality-transferred knowledge was learned during training.

To realize modalities transfer, we need to tackle two problems: i) frame lengths of audio and visual modalities are unaligned due to their different sampling rates; ii) the transferred auditory/visual speech may import new knowledge and feature noise from the other modal feature space at the same time. Other recent works on multi-modal correlation studies on text, vision, and speech include linear, bilinear projection \cite{gao2016compact}, gate network \cite{qian2021audio}, and cross-attention \cite{cheng2020look}. Among these methods, cross-attention is an elegant approach for aligning two temporal sequences of different lengths. 

In this work, we study the cross-modal correlation between auditory and visual speech. We refer to this task as cross-modal co-learning. The key challenge of cross-modal co-learning is modality transfer \cite{rahate2022multimodal}, i.e., modeling one modality aided by exploiting knowledge from another modality using the transfer of knowledge between modalities. We propose a MaxFormer, which manages the cross-modal temporal alignment and knowledge transfer. Furthermore, we validate our method on the compiled audio-visual lip (AVL) database. Code and dataset will be released upon publication\footnote{\url{https://github.com/DanielMengLiu/AudioVisualLip}}.

\begin{figure*}[htbp]
	\centering
	\includegraphics[width=0.90\linewidth]{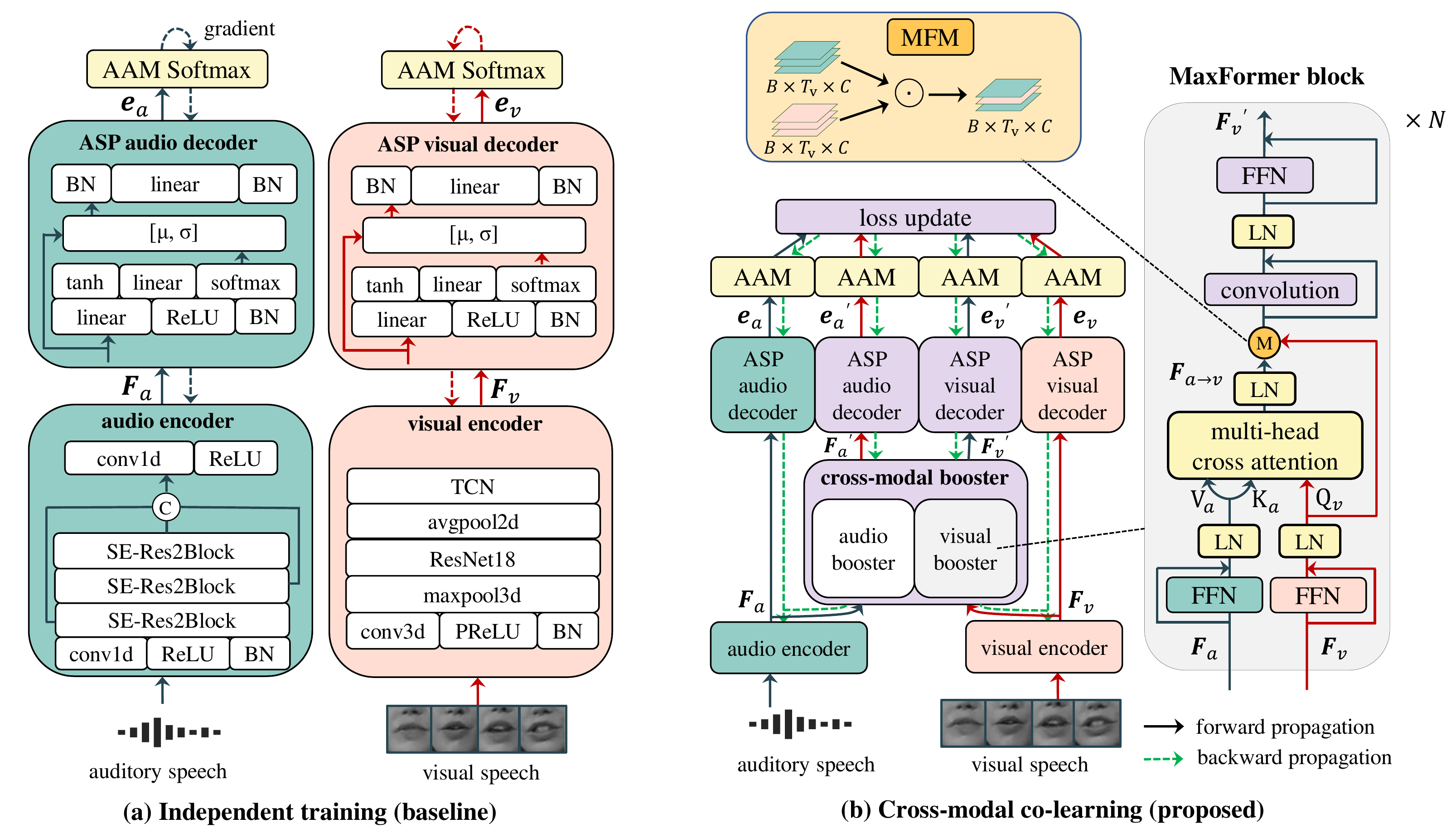}
	\caption{Audio-visual lip biometrics baseline and the proposed cross-modal co-learning framework. }
	\label{fig:AVlipbiometrics}
\end{figure*} 

\section{Cross-modal Speech Co-learning}

Figure \ref{fig:AVlipbiometrics} illustrates our baseline and the proposed cross-modal auditory/visual speech co-learning systems. The baseline systems use two independent branches to learn the audio and visual representation without inter-modality interaction. Specifically, the audio-only encoder involves an ECAPA-TDNN \cite{desplanques2020ecapa} structure, while the visual-only encoder uses an MCNN \cite{liu2021deeplip}. Based on an audio-visual pseudo-siamese structure, we construct the cross-modal co-learning network, illustrated in Fig. \ref{fig:AVlipbiometrics}(b). The whole network consists of two encoders, two cross-modal boosters, and four decoders.

\subsection{Encoders and decoders}

In the audio encoder, the extracted fbank feature is processed by a series of conv1d, ReLU, and batch normalization units. The output frame-level features are concatenated after three SE-Res2Blocks. We use a conv1d with ReLU activation function to obtain a low-dimensional feature map $\boldsymbol{F}_a\in \mathbb{R}^{{C_a}{\times}{T_a}}$, where ${C_a}$ and ${T_a}$ corresponds to channel and frame size, respectively.

In the visual encoder, the 
gray-scale lip sequence $\boldsymbol{L} \in \mathbb{R}^{H{\times}W{\times}T_v}$ is processed by a front-end 3D convolution to extract short-term temporal-spatial visual features, where $H$, $W$ and $T_v$ denote height, width and frame size, respectively. Then a vanilla 18-layer residual network is used to extract spatial features. After that, we apply average pooling to obtain the visual embeddings on each channel. Finally, a temporal convolutional network (TCN) \cite{ma2021towards} captures long-term temporal lip movements $\boldsymbol{F}_v \in \mathbb{R}^{{C_v}{\times}{T_v}}$, where $C_v$ and $T_v$ denote the channel and frame size, respectively. 

All the decoders in this paper apply the attentive statistics pooling (ASP) \cite{okabe2018attentive} decoder, which aggregates the attentive statistics pooling and an embedding affine layer. We use attention without global context \cite{desplanques2020ecapa} for the calculation of ASP. After each decoder, we obtain the output embedding $\boldsymbol{e}$.


\subsection{Cross-modal booster}

To leverage the information from the other modality and fully utilize the concurrency of the audio and visual modalities, we propose the cross-modal boosters. Since the audio and visual booster are symmetric in our co-learning network and share a similar model structure, we take the visual MaxFormer block as an example. Each cross-modal booster contains a stack of MaxFormer blocks, illustrated in Fig. \ref{fig:AVlipbiometrics}(b).


Inside each MaxFormer block, the frame-level audio feature $\boldsymbol{F}_a$ and visual feature $\boldsymbol{F}_v$ are first affine transformed to $d$-dimensional features. After a feed-forward network (FFN) and a layer normalization (LN) block, we construct the input features to a cross-modal triplet (query $\boldsymbol{Q}_v$, key $\boldsymbol{K}_a$, and value $\boldsymbol{V}_a$). Then, the multi-head cross-modal attention calculates the correlation across modalities and aligns the key to the query.  The transferred feature $\boldsymbol{F}_{a \rightarrow v}^{i}$ derived from the $i$-th single head ($i\in\{1,\cdots,m\}$, $m$ demotes the number of heads) is detailed as follows:
\begin{equation}
\boldsymbol{F}_{a \rightarrow v}^{i}=\operatorname{softmax}\left(\frac{(\boldsymbol{Q}_v \boldsymbol{W}_{Q_v}^{i}) (\boldsymbol{K}_a \boldsymbol{W}_{K_a}^{i})^{T}}{\sqrt{d}}\right) \boldsymbol{V}_a \boldsymbol{W}_{V_a}^{i},
\end{equation}
where $\boldsymbol{W}^i$ is the corresponding single head weight. Then the outputs from all attention heads are concatenated and the transferred modality feature $\boldsymbol{F}_{a \rightarrow v}$ is obtained as follows:

\begin{equation}
\boldsymbol{F}_{a \rightarrow v}=[\boldsymbol{F}_{a \rightarrow v}^{1},\ldots, \boldsymbol{F}_{a \rightarrow v}^{i} \ldots,\boldsymbol{F}_{a \rightarrow v}^{m}]\boldsymbol{W}_{a \rightarrow v} 
\end{equation}


\begin{table*}[!ht]
\caption{Dataset statistics of compiled audio-visual lip database (M: male, F: female).} 
\label{tab:datasets}
\centering
\begin{tabular}{llrrlcl}
\toprule
~ & \textbf{Subset} & \textbf{\#Speakers} & \textbf{\#Utterances} & \textbf{Scenario} & \textbf{Video Resolution} & \textbf{Source Task} \\
\midrule
LRSLip3-Train & Train & 4,004 & 31,982 & TED & 224x224 & Lipreading \\ 
LRSLip3-Test  & Test & 412 & 1,321 & TED & 224x224 & Lipreading \\ 
LomGridLip & Test & 24M, 30F & 5,400 & Lab & 720x480 & Lipreading \\ 
GridLip & Test & 18M, 16F & 34,000 & Lab & 360x288 & Lipreading \\ 
\midrule
VoxLip2-Dev & Train & 3,656M, 2,338F & 1,092,007 & YouTube & 224x224 & Speaker Verification \\ 
VoxLip1-Test \cite{nagrani2017voxceleb} & Test & 16M, 13F & 4,162 & YouTube & 224x224 & Speaker Verification \\ 
\bottomrule
\end{tabular}
\end{table*}

To solve the second issue highlighted in Section 1,  we introduce the max feature map (MFM) operation \cite{yang2017max}. The MFM plays the role of local feature selection in Maxformer. It selects the optimal features learned from different modalities at different locations. In the case of backpropagation, it induces a gradient of 0 and 1 to inhibit or activate neurons. The MFM operation can obtain more competitive nodes by activating the maximum value of the feature map, thus reducing feature noise and distortion. The transformed visual speech $\boldsymbol{F}_{a \rightarrow v}$ and the original visual speech compete to compose the enhanced visual speech $\boldsymbol{F}_v^{'}$, shown as follows:

\begin{equation}
\boldsymbol{F}_{v}^{'} = \mathcal{G}_{\theta_{2}}(max(\mathcal{F}_{\theta_{1}}(\boldsymbol{F}_{v}),\boldsymbol{F}_{a \rightarrow v})),
\end{equation}
where $\mathcal{F}_{\theta_{1}}(\cdot)$ is parameter function of layers before the MFM module and $\mathcal{G}_{\theta_{2}}(\cdot)$ corresponds to layers after the MFM module, including convolution \cite{zhang2022mfa}, LN and FFN modules.
\subsection{Cross-modal co-learning loss}
During the training stage, our co-learning loss $\mathcal{L}_{co}$ involves four additive angular margin softmax (AAMSoftmax) \cite{deng2019arcface} loss functions with equal weights, as follows:
\begin{equation}
\mathcal{L}_{co} = \mathcal{L}_{a} + \mathcal{L}_{v} + \mathcal{L}_{a}^{'} + \mathcal{L}_{v}^{'},
\end{equation}
where $\mathcal{L}_{a}$ and $\mathcal{L}_{v}$ represent the loss of the audio and visual speaker embeddings, respectively. $\mathcal{L}_{a}^{'}$ and $\mathcal{L}_{v}^{'}$ denote the loss of the transferred audio and visual embeddings, respectively.

\subsection{Modality fusion}
The auditory speech score $s_a$, visual speech score $s_v$, transferred auditory speech score $s_a^{'}$ and transferred visual speech score $s_v^{'}$ follow two symmetric fusion strategies. The audio- and visual-driven fusion are calculated as follows:
\begin{equation}
s_{a-driven} = 0.5\cdot{s_{a}} +  0.25\cdot{s_{v}} + 0.25\cdot{s_{v}^{'}}
\end{equation}
\begin{equation}
s_{v-driven} = 0.5\cdot{s_{v}} +  0.25\cdot{s_{a}} +  0.25\cdot{s_{a}^{'}} 
\end{equation}
We set the weights considering the contribution of primary, auxiliary, and transferred modalities. The weights could  be determined according to the specific scenario.

\section{Experiments}

\subsection{Data description}

\begin{table}[b]
\caption{Statistics of constructed trial pairs.}
\label{tab:trials}
\centering
\begin{tabular}{lrrr}
\toprule
\textbf{Trial Name} & \textbf{\#Pairs} & \textbf{\#Positive} & \textbf{\#Negative}\\
\midrule
LRSLip3-O & 13,064 & 3,064 & 10,000 \\ 
LomGridLip-O & 20,000 & 4,000 & 16,000 \\ 
GridLip-O & 20,000 & 4,000 & 16,000 \\ 
Vox1-O-29 & 29,690 & 15,057 & 14,633 \\ 
Vox1-O & 37,611 & 18,802 & 18,809 \\ 
\bottomrule
\end{tabular}
\end{table}

As shown in Table \ref{tab:datasets}, we compile an audio-visual lip biometrics dataset from LRS3 \cite{afouras2018lrs3}, LombardGrid \cite{alghamdi2018corpus}, Grid \cite{cooke2006audio}, VoxCeleb1 \cite{nagrani2017voxceleb}, VoxCeleb2 \cite{chung2018voxceleb2}. Almost all the subsets have a similar gender distribution. The LRS3Lip3 and VoxLip subsets were collected from the Internet, covering a wide range of accents, ages, ethnicities, and languages, while LomGridLip and GridLip were collected with lab cameras that the volunteers were stuff and students aging from 18 to 30 years old. The LRS3Lip3 and VoxLip subsets have poor lip resolution due to a transmission loss of the Internet. Verification trials are drawn from the cross-pairs of all test utterances randomly, as shown in Table \ref{tab:trials}. Since some videos in VoxCeleb1 test list were not available from YouTube, we could only download 29/40 test speakers, with 29,690/37,611 selected pairs. We refer to the new trial as Vox1-O-29.

\begin{table*}[htbp]
\caption{Performance comparison between the baseline and cross-modal co-learning systems on various audio-visual lip test sets.}
\label{tab:lrs3}
\centering
\resizebox{\textwidth}{!}{%
\begin{tabular}{|c|l|l|llll|llll|}
\hline
\multicolumn{1}{|l|}{\multirow{2}{*}{}} &
  \multirow{2}{*}{System} &
  \multirow{2}{*}{param} &
  \multicolumn{2}{c|}{LRSLip3-O} &
  \multicolumn{2}{c|}{Vox1Lip-O-29} &
  \multicolumn{2}{c|}{LomGridLip-O} &
  \multicolumn{2}{c|}{GridLip-O} \\ \cline{4-11} 
\multicolumn{1}{|l|}{} &
   &
   &
  \multicolumn{1}{l|}{EER} &
  \multicolumn{1}{l|}{minDCF} &
  \multicolumn{1}{l|}{EER} &
  minDCF &
  \multicolumn{1}{l|}{EER} &
  \multicolumn{1}{l|}{minDCF} &
  \multicolumn{1}{l|}{EER} &
  minDCF \\ \hline
\multirow{3}{*}{baseline} &
  audio-only &
  6.19M &
  \multicolumn{1}{l|}{3.92\%} &
  \multicolumn{1}{l|}{0.2134} &
  \multicolumn{1}{l|}{15.18\%} &
  0.6269 &
  \multicolumn{1}{l|}{5.28\%} &
  \multicolumn{1}{l|}{0.2989} &
  \multicolumn{1}{l|}{4.95\%} &
  0.3291 \\ \cline{2-11} 
 &
  visual-only &
  6.78M &
  \multicolumn{1}{l|}{4.22\%} &
  \multicolumn{1}{l|}{0.1756} &
  \multicolumn{1}{l|}{18.08\%} &
  0.6850 &
  \multicolumn{1}{l|}{2.01\%} &
  \multicolumn{1}{l|}{0.0967} &
  \multicolumn{1}{l|}{2.13\%} &
  0.1241 \\ \cline{2-11} 
 &
  avfusion &
   &
  \multicolumn{1}{l|}{\textbf{1.86\%}} &
  \multicolumn{1}{l|}{\textbf{0.0647}} &
  \multicolumn{1}{l|}{\textbf{11.03\%}} &
  \textbf{0.5423} &
  \multicolumn{1}{l|}{\textbf{0.70\%}} &
  \multicolumn{1}{l|}{\textbf{0.0514}} &
  \multicolumn{1}{l|}{\textbf{0.80\%}} &
  \textbf{0.0471} \\ \hline
\multirow{6}{*}{\begin{tabular}[c]{@{}c@{}}co-learn\\ from\\ scratch\end{tabular}} &
  audio &
  6.19M &
  \multicolumn{1}{l|}{3.85\%} &
  \multicolumn{1}{l|}{0.2194} &
  \multicolumn{1}{l|}{15.65\%} &
  0.6270 &
  \multicolumn{1}{l|}{5.38\%} &
  \multicolumn{1}{l|}{0.3086} &
  \multicolumn{1}{l|}{5.00\%} &
  0.3590 \\ \cline{2-11} 
 &
  visual &
  6.78M &
  \multicolumn{1}{l|}{4.08\%} &
  \multicolumn{1}{l|}{0.1526} &
  \multicolumn{1}{l|}{18.20\%} &
  0.6756 &
  \multicolumn{1}{l|}{1.80\%} &
  \multicolumn{1}{l|}{0.1148} &
  \multicolumn{1}{l|}{2.71\%} &
  0.1596 \\ \cline{2-11} 
 &
  audio-transferred &
  0.89M &
  \multicolumn{1}{l|}{1.73\%} &
  \multicolumn{1}{l|}{0.0777} &
  \multicolumn{1}{l|}{15.61\%} &
  0.6350 &
  \multicolumn{1}{l|}{1.68\%} &
  \multicolumn{1}{l|}{0.1135} &
  \multicolumn{1}{l|}{1.40\%} &
  0.0926 \\ \cline{2-11} 
 &
  visual-transferred &
  0.89M &
  \multicolumn{1}{l|}{2.16\%} &
  \multicolumn{1}{l|}{0.1019} &
  \multicolumn{1}{l|}{15.79\%} &
  0.6687 &
  \multicolumn{1}{l|}{1.20\%} &
  \multicolumn{1}{l|}{0.0839} &
  \multicolumn{1}{l|}{1.45\%} &
  0.1006 \\ \cline{2-11} 
 &
  audio-driven fusion &
  \multirow{2}{*}{} &
  \multicolumn{1}{l|}{\textbf{1.37\%}} &
  \multicolumn{1}{l|}{\textbf{0.0460}} &
  \multicolumn{1}{l|}{\textbf{11.00\%}} &
  \textbf{0.5384} &
  \multicolumn{4}{c|}{-} \\ \cline{2-2} \cline{4-11} 
 &
  visual-driven fusion &
   &
  \multicolumn{4}{c|}{-} &
  \multicolumn{1}{l|}{\textbf{0.83\%}} &
  \multicolumn{1}{l|}{\textbf{0.0414}} &
  \multicolumn{1}{l|}{\textbf{0.85\%}} &
  \textbf{0.0538} \\ \hline
\multirow{6}{*}{\begin{tabular}[c]{@{}c@{}}co-learn\\ with \\ pretrained\end{tabular}} &
  audio &
  6.19M &
  \multicolumn{1}{l|}{4.16\%} &
  \multicolumn{1}{l|}{0.2220} &
  \multicolumn{1}{l|}{15.12\%} &
  0.6215 &
  \multicolumn{1}{l|}{5.54\%} &
  \multicolumn{1}{l|}{0.3242} &
  \multicolumn{1}{l|}{5.13\%} &
  0.3796 \\ \cline{2-11} 
 &
  visual &
  6.78M &
  \multicolumn{1}{l|}{3.92\%} &
  \multicolumn{1}{l|}{0.1607} &
  \multicolumn{1}{l|}{17.62\%} &
  0.6641 &
  \multicolumn{1}{l|}{1.82\%} &
  \multicolumn{1}{l|}{0.0794} &
  \multicolumn{1}{l|}{1.80\%} &
  0.1184 \\ \cline{2-11} 
 &
  audio-transferred &
  0.89M &
  \multicolumn{1}{l|}{2.55\%} &
  \multicolumn{1}{l|}{0.0994} &
  \multicolumn{1}{l|}{16.32\%} &
  0.6225 &
  \multicolumn{1}{l|}{2.09\%} &
  \multicolumn{1}{l|}{0.0994} &
  \multicolumn{1}{l|}{0.80\%} &
  0.0568 \\ \cline{2-11} 
 &
  visual-transferred &
  0.89M &
  \multicolumn{1}{l|}{3.64\%} &
  \multicolumn{1}{l|}{0.1461} &
  \multicolumn{1}{l|}{17.52\%} &
  0.6483 &
  \multicolumn{1}{l|}{1.86\%} &
  \multicolumn{1}{l|}{0.1011} &
  \multicolumn{1}{l|}{2.21\%} &
  0.1395 \\ \cline{2-11} 
 &
   avfusion &
   & 
  \multicolumn{1}{l|}{1.70\%} &
  \multicolumn{1}{l|}{0.0612} &
  \multicolumn{1}{l|}{10.77\%} &
  0.5246 &
  \multicolumn{1}{l|}{0.82\%} &
  \multicolumn{1}{l|}{0.0519} &
  \multicolumn{1}{l|}{0.74\%} &
  0.0464 \\ \cline{2-11} 
 &
  audio-driven fusion &
  \multirow{2}{*}{} &
  \multicolumn{1}{l|}{\textbf{1.50\%}} &
  \multicolumn{1}{l|}{\textbf{0.0504}} &
  \multicolumn{1}{l|}{\textbf{10.27\%}} &
  \textbf{0.5159} &
  \multicolumn{4}{c|}{-} \\ \cline{2-2} \cline{4-11} 
 &
  visual-driven fusion &
   &
  \multicolumn{4}{c|}{-} &
  \multicolumn{1}{l|}{\textbf{0.90\%}} &
  \multicolumn{1}{l|}{\textbf{0.0407}} &
  \multicolumn{1}{l|}{\textbf{0.45\%}} &
  \textbf{0.0311} \\ \hline
\end{tabular}%
}
\end{table*}

\subsection{Experimental setup}
We train the network for 40 epochs with a batch size equal to 128, a multi-step learning scheduler (milestones are 10 and 15, gamma is 0.1), an initial learning rate of 0.001, and a weight decay of 1e-7, with the Adam optimizer. The scale and margin of AAMSoftmax are set to 30 and 0.2, respectively. During the training stage, 2 seconds of audio-visual clips are used. For visual clips, we use 50$\times$96$\times$96 pixels and a random horizontal flip for augmentation. For audio clips, we extract 80-dimensional fbanks. Within an epoch, we apply the equal possibility of choosing clean, noisy (generated using MUSAN \cite{musan} and RIRs \cite{rirs}) and spec augmentation \cite{park2019specaugment} samples. 

The audio encoder uses a 512-dimensional channel. The visual encoder has two layers of TCN with a kernel size of 5. Each cross-modal booster uses three blocks of MaxFormer, and $d$ is set to 128. All decoders use a 192-dimensional output embedding.
Performance will be measured by providing the Equal Error Rate (EER) and the minimum normalized detection cost function (MinDCF)  \cite{desplanques2020ecapa} with $P_{target}{=}10^{-2}$   and $C_{FA}{=}C_{Miss}{=}1$.

\subsection{Training on LRSLip3}

The models are all trained on the LRSLip3 training set. For the baseline, the audio-only and visual-only systems are trained independently, and then their embeddings are fused. As for our proposed cross-modal co-learning method, we train the network both from scratch and with pretrained audio- and visual-only models.

Table \ref{tab:lrs3} compares the performance between the baseline and the proposed co-learning systems. Our proposed system can reach EERs of 1.37\%, 10.27\%, 0.83\%, and 0.45\% on LRSLip3-O, Vox1-O-29, LomGridLip-O, and GridLip-O test sets, and the minDCFs reduce by 28.9\%, 4.9\%, 20.8\%, and 34.0\%, respectively. Generally speaking, the transferred audio and visual speech achieved improvement compared with the corresponding unimodal systems. 
The fusion system further improves the performance, which indicates that our proposed cross-modal boosters have learned new knowledge. However, we notice that the performance of our co-learning model (from scratch) declines a bit on GridLip-O, which may be due to over-fitting to the training set.

Another interesting finding is that the visual speech performs better than the auditory speech on the LomGridLip and GridLip sets, and vice versa. The LRSLip3 and VoxLip sets were collected from the Internet with poor visual resolution, and the LomGridLip and GridLip sets were collected using lab cameras with high visual resolution. Experimental results indicate the benefit of choosing visual- or audio-driven fusion according to data conditions. With a high-resolution camera, we can utilize visual speech for near-field audio-visual authentication, e.g., with the mobile phone or at the bank counter.

\begin{figure}[htbp]
	\centering
	\includegraphics[width=0.8\linewidth]{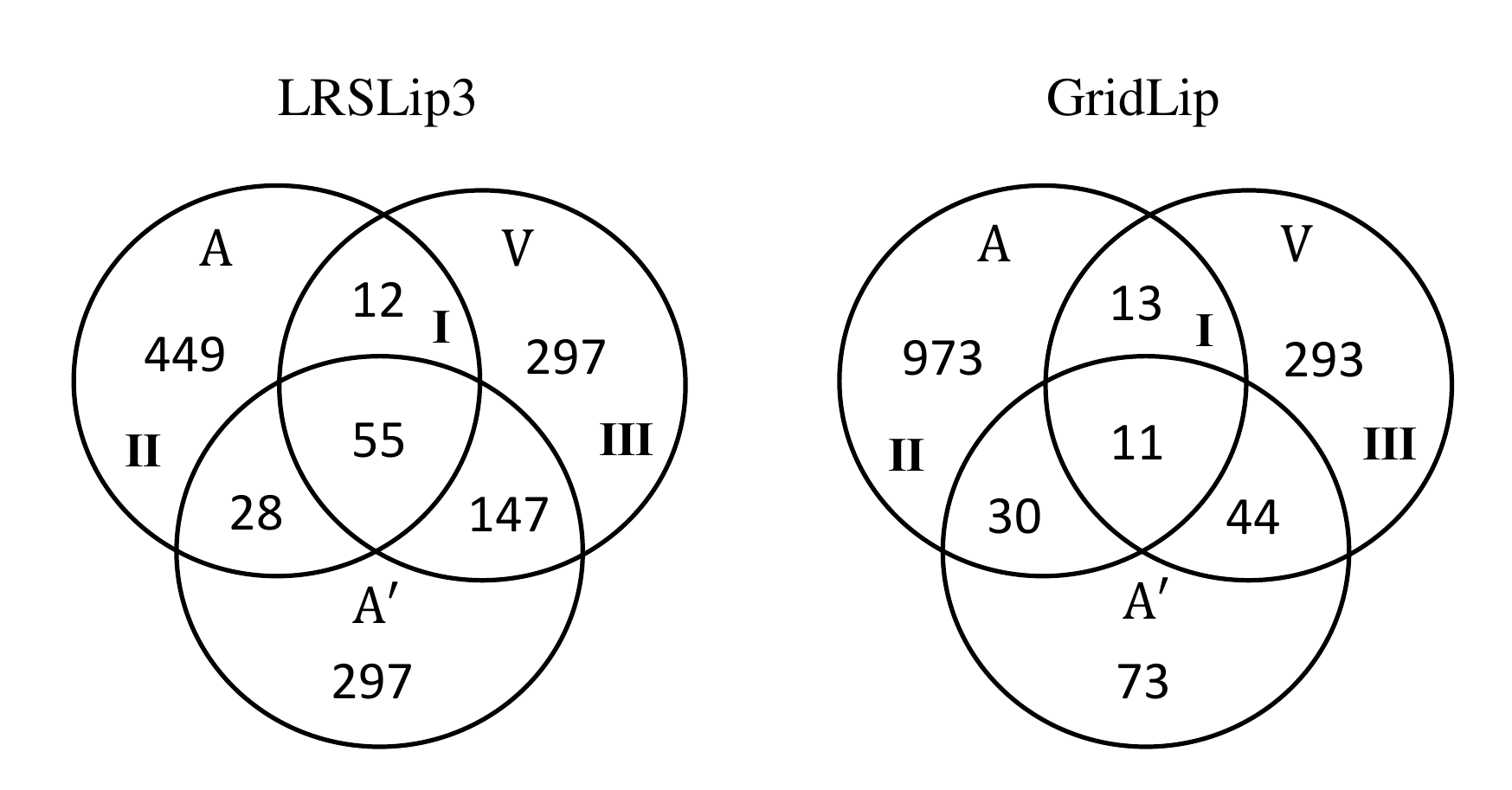}
	\caption{Analysis on wrong verification predictions using the co-learning model (A: auditory speech, V: visual speech, A': transferred auditory speech).}
	\label{fig:analysis}
\end{figure} 

Furthermore, we analyze the verification predictions on the LRSLip3 and GridLip sets using the co-learning method. As shown in Fig. \ref{fig:analysis} (left), area \uppercase\expandafter{\romannumeral+1} represents twelve verification pairs that are wrongly predicted by both audio and visual speeches but correctly predicted by transferred auditory speech. It is the new knowledge learned via modality transfer and does not exist in the original audio or visual feature space. Area \uppercase\expandafter{\romannumeral+2} and \uppercase\expandafter{\romannumeral+3} contain 
new knowledge that benefits for fusion systems.

\subsection{Training on VoxLip}

As shown in Table \ref{tab:lrs3}, the cross-modal booster seems to have a mismatch on the VoxLip1 set, in which both biometric modalities have poor performance; thus, correlation is hard to learn. Furthermore, the train set of LRSLip3 has limited utterances which are challenging to train a robust model for the VoxLip1 set. Therefore, we train on the VoxLip2 set and co-learn with the pretrained model.

\begin{table}[htbp]
\caption{Performance comparison between the baseline and co-learning systems on the VoxCeleb1 test sets (w/o AS-norm).}
\label{tab:vox}
\centering
\resizebox{0.48\textwidth}{!}{%
\begin{tabular}{lllcl}
\hline
\multirow{2}{*}{System} & \multicolumn{2}{l}{Vox1-O-29} & \multicolumn{2}{l}{Vox1-O}          \\ \cline{2-5} 
                        & EER           & minDCF        & \multicolumn{1}{l}{EER}    & minDCF \\ \hline
audio-only              & 1.21\%        & 0.0971        & \multicolumn{1}{l}{1.16\%} & 0.0877 \\ 
visual-only             & 6.75\%        & 0.2189        & \multicolumn{2}{c}{-}               \\ 
fusion                  & 1.14\%        & 0.0759        & \multicolumn{2}{c}{-}               \\ 
AV-Hubert-L(AV) \cite{shi2022learning}       & -        & -        & 0.84\% & -               \\ \hline
audio-transferred       & 2.89\%        & 0.1500        & \multicolumn{2}{c}{-}               \\ 
visual-transferred      & 5.33\%        & 0.1919        & \multicolumn{2}{c}{-}               \\ 
audio-driven fusion     & \textbf{0.76\%}        & \textbf{0.0559}        & \multicolumn{2}{c}{-}               \\ \hline
\end{tabular}
}
\end{table}

As shown in Table \ref{tab:vox}, without AS-norm, the baseline fusion system achieves an EER of 1.14\% and a minDCF of 0.0759 on Vox1-O-29. With the cross-modal knowledge, the system performance could further reach an EER of 0.76\% with a relative reduction of 33.3\%. 

\section{Conclusion and Future Work}
This paper investigated the correlation between speech and lip motion and proposed a cross-modal speech co-learning paradigm. A cross-modal booster structure has been presented to learn the modality-transformed correlation. Our primary contribution is the cross-modal co-learning paradigm that realizes knowledge transfer between auditory and visual speech. We generated enhanced features via an MFM-embedded MaxFormer. Experiments on multiple test sets revealed the potential application scenario of the proposed audio-visual speaker recognition using lips. In the future, we will continue working on text-dependent audio-visual speaker verification using lips.

\section{Acknowledgements}
This work was supported by the National Key R\&D Program of China under Grant 2018YFB1305200, the National Natural Science Foundation of China under Grant 61771333 and the Tianjin Municipal Science and Technology Project under Grant 18ZXZNGX00330.

\vfill\pagebreak
\bibliographystyle{IEEEtran}
\begin{spacing}{0.1}
\ninept
\bibliography{mybib.bib}
\end{spacing}

\end{document}